%% file: paper5.tex
\documentstyle[aps,prl,preprint]{revtex}

\begin{document}


\title{The Effect of Resonances on Diffusive Scattering}
\author{B. Elattari\cite{brahim}, V. Kagalovsky, and H.A. Weidenm\"uller}
\address{Max-Planck-Institut f\"ur Kernphysik, 69029 Heidelberg, Germany}
\date{\today}
\maketitle

\begin{abstract}The presence of resonances modifies the passage of light or
of electrons through a disordered medium. We generalize random matrix theory
to account for this effect. Using supersymmetry, we calculate analytically
the mean density of states, and the effective Lagrangean of the generating
functional for the two--point function. We show that the diffusion constant
scales with the effective mean level spacing. The latter exhibits a
resonance dip. These facts allow us to interpret experimental results
on light scattering for different concentrations of resonant scatterers.
\end{abstract}

\section{Introduction and Motivation}During the last decade, many 
phenomena related to the passage of waves through a disordered medium
have been intensely studied. This holds both for electrons, i.e. for
the amplitude waves of the Schr\"odinger equation \cite{review}, and
for light, i.e. for classical electromagnetic waves
\cite{reviewlight}. As examples, we mention 
conductance fluctuations in mesoscopic probes, the analogous effect
of speckle patterns in the transmission of light, and weak and strong
localization effects including the enhanced backscattering of electrons
or light from a disordered medium. For electrons, disorder is typically
caused by impurities. For light, disorder is often produced
artificially: A powder is immersed into some liquid. Random scattering
occurs if the indices of refraction of powder and liquid are sufficiently
different.  

The present paper has been motivated by an experiment performed several
years ago \cite{Alb91} on the scattering of light by a disordered
medium. In this experiment, the powder used (TiO$_2$) consisted of 
grains with a size distribution centered around a diameter of $220$
nm. For such granules, a Mie resonance occurs close to the wavelength 
$\lambda \sim 630$ nm of the laser light used in the experiment. This 
led to a resonance enhancement of the diffusive scattering. The effects
of such enhancement were seen by comparing the diffusion constant 
${\cal D}$ (determined from the intensity autocorrelation function versus
frequency of the transmitted light), with the transport mean free path $l$
(determined either from weak localization, i.e. from enhanced
backscattering, or from the dependence of the transmitted intensity on
the length $L$ of the disordered sample). For the quasi one--dimensional
geometry of the present experiment, one expects ${\cal D}$ and $l$ to obey
the relation ${\cal D} = (1/3) \ v_E l$, with $v_E$ the energy transport
velocity through the disordered medium. In the present case, the ratio
$3 {\cal D} / l$ yielded a value $v_E = (5 \pm 1) \ 10^7$  m s$^{-1}$ which
is about an order of magnitude smaller than the phase velocity. This
surprising result has been understood both for small concentration 
\cite{Alb91,KoganK,Cwi,Kroha,Barab,vanT93,vanTE93} and, more recently,
also for strong concentration \cite{Bus95,Bus96} of the scatterers. (In
the latter case, the resonant structure in the frequency dependence of
${\cal D}$ disappears, and an overall decrease of ${\cal D}$ is observed
\cite{Gar}). Qualitatively speaking, the transport velocity is reduced
because on its way through the medium, the energy is stored for some time
in the Mie resonances.

The present paper goes beyond previous theoretical work in two ways. 
First, we present a unified theoretical framework in which both
average and fluctuation properties of the transmitted and reflected
waves can be calculated on the same footing. This is in contrast to
approaches \cite{Alb91,Bus95,Bus96} which use the Bethe--Salpeter
equation \cite{Alb91} or a mean--field approximation \cite{Bus95,Bus96}
for the calculation of the transport velocity $v_E$, and a diagrammatic
impurity perturbation expansion for the intensity autocorrelation function.
Second, we identify the energy dependence of the mean level density
$\rho(E)$ as the culprit for the observed deviation from standard behavior.
A simple argument which yields the same result as the analytical derivation
and illuminates the physical content may be helpful at this point. In the
case of electrons \cite{review}, the Thouless formula $g = E_c \rho(E)$
connects the average conductance $g$ with the Thouless energy $E_c = \hbar
{\cal D} / L^2$ and the mean level density. The presence of numerous
resonances with equal resonance energies $E_1$ leads to a local
Breit--Wigner--shaped increase of $\rho(E)$ near $E_1$. Since $g$ is
not affected by the presence of the resonances, the Thouless formula
implies that $E_c$ and, hence, the diffusion constant ${\cal D}$ have
a Breit--Wigner--like dip near $E_1$. In Refs. \cite{ourprl,ourpre} we
have shown that within the framework of the supersymmetry formalism, the 
effective Lagrangeans for electrons and for classical light waves coincide.
Hence, the argument just given applies likewise to scattering of light.
With increasing concentration of scatterers, the dip in ${\cal D}$ widens
and becomes less deep. Eventually, this results in an overall decrease of 
${\cal D}$ over a wide frequency interval. We emphasize that in the context
of the supersymmetry approach, the relation ${\cal D} = (1/3) v_E l$ is not
used explicitly. It is replaced by the Thouless relation.

A summary of our results has been given in Ref.~\cite{ourprl}. First
results for electrons were published in Ref.~\cite{our}. In the present
paper we give a detailed derivation of our results, paying special
attention to the necessary modifications of random matrix theory, and
of the standard supersymmetry formalism. For classical waves, the
latter is introduced in Ref.~\cite{ourpre}.

The random matrix model with resonances is introduced in section \ref{M}.
In section \ref{S}, a particularly simple case is presented which paves 
the way for the full results given in section \ref{cond}. Section
\ref{concl} contains the conclusions. We use Efetov's supersymmetric
generating functional \cite{Efe83} in the version of ref. \cite{Ver85}.

\section{The model}
\label{M}

In Ref. \cite{our}, we have given a detailed physical justification
for the choice of the random matrix model used to describe diffusive
scattering in the presence of resonances. The arguments presented there
apply to the case of electrons. Meanwhile, we have shown that the
non--linear sigma models for electrons and classical waves are identical,
save for the source terms \cite{ourprl,ourpre}. For these reasons, we
keep our presentation brief. In this section we always refer to
electrons. It must be understood that all arguments apply equally to
light (classical waves). 

In the {\it absence} of Mie resonances, a quasi one--dimensional
disordered system of length $L$ for diffusive scattering of electrons
is modeled \cite{Iid90} as consisting of many slices of length $l_0 
\ll L$ each. Within each slice, the Hamiltonian is modeled as a
member of the GOE, the random--matrix ensemble with orthogonal
symmetry. Neighboring slices are coupled by Gaussian--distributed
uncorrelated random matrix elements. The strength of this coupling
defines the diffusion constant. The first and the last slice are
coupled to the channels, i.e., the asymptotic states in the
leads. Elements of the scattering matrix connecting incident and
outgoing channels in both leads define the transmission through the
disordered  region and, thus, the conductance.  

To account for the presence of Mie resonances, we retain the idea of 
a division of the quasi one--dimensional system into slices but modify
the form of the Hamiltonian $\bf H$. Within each slice, $\bf H$ is a
matrix of dimension $N + m$ where $m$ is the number of Mie scatterers
within each slice. This number is determined by the concentration of
scatterers, and by the linear dimensions of the slice. Thus, $\bf H$
has the form
\begin{equation}
{\bf H}=\left( \begin{array}{cc} \begin{array}[b]{c}
H_{\rm GOE} \end{array} & V \\
V^T & E_1\times I_m+H_{\rm res}
\end{array}
\right)
\label{ham}
\end{equation}
where $T$ denotes the transpose. Here, $H_{\rm GOE}$ is a matrix of
dimension $N$ belonging to the Gaussian orthogonal ensemble (GOE) of
random matrices, $E_1 \times I_m$ is an $m$--dimensional diagonal matrix
with elements $E_1 = \hbar \omega_1$ corresponding to the presence of
$m$ Mie scatterers with equal resonance frequencies $\omega_1$, $I_m$ is
the $m$--dimensional unit matrix, $H_{\rm res}$ is a random matrix of
dimension $m$, also a member of the GOE which describes the coupling
between resonances, and $V$ is a rectangular matrix which couples the 
$m$ resonances to $H_{\rm GOE}$. We assume the matrices $V$, $H_{\rm res}$,
and $H_{\rm GOE}$ to be statistically uncorrelated. The variance of the
matrix elements of $H_{\rm GOE}$ is chosen as usual: The ensemble average
of $(H_{\rm GOE})_{\mu \nu} (H_{\rm GOE})_{\mu' \nu'}$ is given by
$(\lambda^2/N) (\delta_{\mu, \mu'} \delta_{\nu, \nu'} + \delta_{\mu,
\nu'} \delta_{\nu, \mu'})$. Here, $\lambda$ has the dimension of an
energy and determines the average level spacing $d_S$ (the same in each
slice). For $m = 0$ and with $\Delta(E) = \sqrt{1 - (E / (2 \lambda))^2}$,
the dependence of $d_S$ on energy $E$ is given by $\pi \lambda / (N
\Delta(E))$ (semicircle law). At the end of the calculation, we take the
limit $N \rightarrow \infty$. The ensemble average of $(H_{\rm res})_{\mu
\nu}(H_{\rm res})_{\mu '\nu '}$ is written analogously as
$(\lambda_{1}^{2}/m) (\delta_{\mu, \mu'} \delta_{\nu, \nu'} +
\delta_{\mu, \nu'} \delta_{\nu, \mu'}) $. The strength factor $\lambda_1$
is chosen in such a way that the interaction between resonances results in
a lifting of the degeneracy which is of the order of the mean level spacing
$d_S(0)$ so that $\lambda_1 \ll \lambda$. Too strong an interaction would
wash out the resonance structure altogether. Finally, and without loss of
generality, the rectangular matrix $V$ can be taken to be diagonal. 
Indeed, $V$ can be written in the form $O_1 V_D O_2^T$ where $V_D$ is
diagonal and rectangular, and where $O_1$ and $O_2$ are orthogonal
matrices of dimensions $N$ and $m$, respectively. Transforming $\bf H$
with the orthogonal matrix $O = $ diag$(O_1,O_2)$, and using the
orthogonal invariance of the GOE to absorb the matrices $O_1$ into
$H_{\rm GOE}$ and $O_2$ into $H_{\rm res}$, respectively, we obtain a
new Hamiltonian of the form of Eq.~(\ref{ham}), but with $V$ replaced
by the diagonal rectangular matrix $V_D$. Since we assume all
scatterers to be identical, we take all diagonal matrix elements of
$V_D$ to be equal. We assume that the diagonal matrix elements $v$ 
scale with $N$ in the same way as those of the matrices $H_{\rm GOE}$
and $H_{\rm res}$, so that $v^2 = \alpha \lambda^2 / N$ where $\alpha$
is independent of $N$. This completes the definition of the model. 

It is useful to relate $\lambda_1$ and $v$ microscopically to the
properties of the Mie resonance, and to the concentration $\mu$ of
scatterers. By construction, the total coupling strength of each of
the $m$ resonant states is on average given by $v^2 + \lambda_1^2$. 
This quantity must, therefore, be proportional to the square of the
matrix element for electromagnetic decay of the Mie resonance. 
Moreover, $\lambda_1$ must grow monotonically with increasing
concentration of scatterers because an ever greater fraction of the
light emitted from one resonance will be absorbed by another one,
rather than scattered randomly. More precisely, consider a granule
with a Mie resonance emitting a spherical wave. This wave hits an
(arbitrary) second granule at distance $s$. For geometrical reasons,
the cross section for absorption by excitation of the Mie resonance in
this second granule is proportional to $s^{-2}$. With changing
concentration, the distance $s$ between any pair of granules changes
as $s^3 \sim m^{-1}$. Hence, the coupling strength between any pair of
granules (which we identify with $\lambda_1^2/m$) is proportional to
$m^{2/3}$. This yields $\lambda_1 \sim m^{5/6}$. This consideration
suggests that the coupling between granules should be dependent upon
their distance, a fact not taken into account in Eq.~(\ref{ham}) where
all resonances are coupled in the same way. Thus, using a random band
matrix instead of the GOE matrix $H_{\rm res}$ may yield a better
simulation of the actual situation. We shall find, however, that the
entire effect of $H_{\rm res}$ consists in widening the peak in the
mean density of states. Therefore, it does not seem worthwhile to go
through the trouble of dealing with a more complicated ensemble. On 
the other hand, this consideration shows why it is justified not to
take into account the interaction of granules located in different
slices. This is the approximation we will use.

\section{One Slice with Coupled Resonances}
\label{S}

For pedagogical reasons we focus attention in this section to the
simplest case of a single slice described by the Hamiltonian $\bf H$
given by Eq.~(\ref{ham}). In Ref.~\cite{our}, we have studied the
case of a single slice with uncoupled resonances ($H_{\rm res}=0$). 
Here, we generalize this study to the case of coupled resonances
($H_{\rm res} \neq 0$). We calculate the mean level density and the
level-level correlation function, using the supersymmetry technique of
refs. \cite{Efe83,Ver85}.

The advanced and retarded Green functions can be written as integrals
over supervectors 
\begin{equation}
G_{\mu\nu}^{\pm}(E\pm\omega)=\mp i\int {\rm d} [\Psi]
S_{\mu}^{1}S^{1}_{\nu}\exp [\pm i {\cal L}(\Psi )] ,
\label{gnn}
\end{equation}
where the Lagrangean is given by
\begin{equation}
{\cal L}=\frac{1}{2}\Psi^{\dagger}\left( E\pm\omega\pm i\eta -{\bf H}
\right)\Psi  .
\label{lag}
\end{equation}
The supervectors $\Psi$ are defined by
\begin{equation}
\Psi^{\dagger}=(S^1,S^2,\chi ,
-\chi^{*})  .
\label{vec}
\end{equation}
The vectors $S^1$ and $S^2$ have $N+m$ ordinary real components, whereas 
the vectors $\chi$ have $N+m$ anticommuting components. We define a
source term $J=$diag $(j,j,-j,-j)$ in graded space, where j is a
matrix of dimension $(N+m)$ with entries $J_{\nu \mu}$, and introduce the 
generating functional
\begin{equation}
Z^{\pm}(E\pm\omega ,J)=
\int {\rm d} [\Psi ] \exp \left[ \pm i ({\cal L} - \frac{1}{2}  
\Psi^{\dagger}J\Psi ) \right]  .
\label{gen}
\end{equation}
The Green functions are obtained as functional derivatives of $Z$ with
respect to $J$ at $J = 0$,
\begin{equation}
G^{\pm}_{\mu\nu}=\frac{1}{4}\frac{\delta Z^{\pm}}{\delta J_{\nu\mu}}  .
\label{gre}
\end{equation}
We use this expression to calculate the average of the product of a
retarded and an advanced Green function taken at different energies,
$<G^{+} G^{-}>$. This quantity (the ``two--point function'') plays an
important role in describing average properties of random systems,
such as the level--level correlation function. Except for the
dimensions of the supervectors $\Psi$ and of the $\sigma$-- and
$Q$--matrices appearing below, and except for the occurrence of
additional frequency variables, the $2k$--point function for any
positive integer $k$ is governed by an effective Lagrangean of the
same type.

The generating functional $Z$ for the two--point function is given by
\begin{equation}
Z(E,\omega ,J)=\int {\rm d} [\Psi ] \exp\left[
\frac{1}{2}i
\Psi^{\dagger}L^{1/2}\left(E+\omega L
+i\eta L-{\bf H} - J \right)L^{1/2}\Psi\right] .
\label{gen1}
\end{equation}
We have introduced two $8(N+m)$--dimensional graded matrices, namely 
$L = $diag$(I,I,-I,-I,I,I,-I,-I)$ and $J = $diag$(j,j,-j,-j,j,j,-j,-j)$.
Here, $I$ denotes the $N+m$--dimensional unit matrix. 
The quantity $\Psi$ now denotes a supervector with $8(N+m)$ components. 
All quantities are given in ``advanced--retarded'' notation (see
Ref.~\cite{Ver85}). We average over the two uncorrelated random
ensembles $H_{\rm GOE}$ and $H_{\rm res}$. The presence of these two
ensembles causes the Hubbard--Stratonovich transformation to involve
two supermatrices $\sigma$ and $\sigma_1$. For the generating
functional, we find
\begin{equation}
Z(E,\omega ,J)=\int {\rm d} [\sigma] \exp \left[
-\frac{N}{4}{\rm trg} \sigma^2 - 
\frac{m}{4}{\rm trg} \sigma_{1}^{2} - 
\frac{1}{2}{\rm Trg} \log (M-\omega L -J)\right] ,
\label{fun2}
\end{equation}
where we define
\begin{equation}
M=\left(\begin{array}{ccc}
E-\lambda\sigma & 
0 & -V_D \\ 0 & E-\lambda\sigma & 0 \\ -V_D & 0 & E-E_1-\lambda_1\sigma_1 
\end{array}\right)
\begin{array}{l} \}m \\ \}N-m \\ \}m \end{array} .
\label{mm}
\end{equation}
Here, trg denotes the trace over an $8\times8$ matrix in graded space,
and Trg the trace over an $8(N+m)\times8(N+m)$ matrix in both graded
and Hilbert space. The symbol ${\rm d} [\sigma]$ stands for the 
differentials of all the independent variables in both $\sigma$ and 
$\sigma_1$. The dimensions of the block matrices in Hilbert
space are indicated in Eq.~(\ref{mm}). After taking the limit $N
\rightarrow \infty$ {\it and keeping $m$ fixed}, variation with
respect to $\sigma$ yields the standard saddle--point equation with a
constant diagonal solution  
\begin{equation}
\sigma_D =\frac{E}{2\lambda} \pm i\Delta ,
\label{zho}
\end{equation}
where $\Delta = \sqrt{1-(E/2\lambda )^2}$. Variation with respect to 
$\sigma_1$ produces a second saddle--point equation
\begin{equation}
\sigma_1 =\frac{\lambda_1}{E-E_1-\lambda_1\sigma_1 -\Gamma\sigma} ,
\label{sad2}
\end{equation}
where $\Gamma =v^2/\lambda$. Eq.~(\ref{sad2}) defines the diagonal
elements of $\sigma_1$ in terms of those of $\sigma$. The saddle--point
manifold is generated by the usual choice of signs in Eq.~(\ref{zho}),
and by transforming the resulting diagonal matrices $\sigma_D$ and
$\sigma_{1D}$ both by the $\it same$ matrix $T$. The latter is defined
in Eq.~(5.28) of Ref.~\cite{Ver85}. The resulting matrices are denoted
by $Q$ and $Q_1$, respectively. 

We integrate over the massive modes and expand the logarithm up to terms 
linear in $\omega$ and $J$, using the identity
\begin{eqnarray}
\exp\left [{\rm trg} \log\left(
\begin{array}{cc} \begin{array}[b]{c}
A \end{array} & B \\
B^T & D
\end{array} \right) \right] = \exp \left [{\rm trg} \log \left(
\begin{array}{cc} \begin{array}[b]{c}
A-B D^{-1} B^T \end{array} & 0 \\
0 & D 
\end{array} \right) \right] 
\label{expo}
\end{eqnarray}
and the fact that traces over terms not containing either $L$ or $J$ vanish.
Introducing the diagonal graded matrix $\Lambda = $diag$(1,1,1,1,-1,-1,-1,-1)$,
we find
\begin{equation}
Z(E,\omega ,J)=\int {\rm d} \mu (t)  \  \exp\left [-\frac{N}{\lambda}{\rm
  trg} [(Q + \frac{m\lambda}{N\lambda_1} Q_1) (\omega\Lambda +J)]
\right] \ . 
\label{simp}
\end{equation}
The symbol ${\rm d} \mu(t)$ denotes the invariant measure, see 
Ref.~\cite{Ver85}. 
 
The expression in Eq.~(\ref{simp}) is formally equivalent to the standard 
one, obtained in the absence of resonances. This is seen by introducing a 
new diagonal supermatrix   
\begin{equation}
\sigma_D^{*}=\sigma_D+\frac{m\lambda}{N\lambda_1}\sigma_{1D},
\label{new}
\end{equation}
and by defining a new $Q$--matrix $Q^{*}$ by
\begin{equation}
Q^{*} = T \sigma^{*} T^{-1}.
\label{qmatrix}
\end{equation}
This leads to 
\begin{equation}
Z(E,\omega ,J)=\int {\rm d} \mu (t) \ exp \left[-\frac{N}{\lambda}{\rm
  trg}[(Q^{*}(\omega \Lambda +J)] \right] .
\label{simp2}
\end{equation}

Because of the presence of the $m$ resonances, the average density of 
states is not given directly in terms of the saddle--point solution 
$\sigma_D$ but is obtained by differentiation of the generating functional. 
This yields 
\begin{equation}
\overline{\rho}=\frac{N}{\lambda\pi}{\rm Im}\sigma_D^{*}=
\frac{N\Delta}{\lambda\pi}+\frac{m}{\lambda_1\pi}
{\rm Im}\sigma_1 \ .
\label{dosob}
\end{equation}
The integral over $\rho$ must be equal to the dimension of 
the Hamiltonian, $\int\rho (E)dE=N+m$. This can be verified by 
a straightforward calculation. To obtain a better understanding of the 
form of the density of states given by Eq. (\ref{dosob}), we consider
the two limits of very small and very large $\lambda_1$. In the limit 
$\lambda_1 \rightarrow \infty$ ($\lambda_1\gg \Gamma$), the solution of
Eq.~(\ref{sad2}) to zeroth order in $\Gamma$ is given by
\begin{equation}
 \sigma_{1D}=\frac{E-E_1}{2\lambda_1}+i\Delta^{'},
\label{sig1}
\end{equation}
where $\Delta^{'}=\sqrt{1-((E-E_1)/2\lambda_1)^2}$. 
For sufficiently large $\lambda_1$, i.e. sufficiently strong coupling 
between the resonances, the coupling of the Hamiltonians $H_{\rm GOE}$ and 
$H_{\rm res}$ can be neglected, and the density of states reduces to the 
sum of two semicircles 
\begin{equation}
\overline{\rho}=\frac{N\Delta}{\lambda\pi}+\frac{m\Delta^{'}}{\lambda_1\pi}.
\label{rhob}
\end{equation}  
We see that the resonant structure is completely washed out. This was to be 
expected. In the opposite limit $\lambda_1 \rightarrow 0$ (uncoupled 
resonances), it is easily seen that we retrieve our earlier result for 
uncoupled resonances \cite{our}
\begin{equation}
\overline{\rho (E)} = \frac{\Delta (E)N}{\lambda\pi}+
m\frac{\Gamma\Delta /\pi}
{(E-E_1-(E/2\lambda )\Gamma )^2+(\Gamma\Delta )^2},
\label{dos1}
\end{equation}
where $\pi \lambda / (N \Delta)$ is the mean level spacing $d_S$ in the 
absence of resonances. The second term in Eq.~(\ref{dos1}) has area $m$, peak 
height $m / (\pi \Gamma \Delta)$, and Lorentzian shape. 
As expected, Eq.~(\ref{dos1}) shows a resonance enhancement of the mean level 
density, centered at the energy $E_1$ of the $m$ resonances. The total width
($\Gamma \Delta$) is determined by the strength $v$ of the coupling to
the random scatterers. We recall the relation $v^2 = \alpha \lambda^2
/ N$ and find  $\Gamma \Delta = \alpha \lambda \Delta / N  = (\alpha /
\pi) d_S$. This shows that $\Gamma$ is of the order of the mean level
spacing $d_S$. For light scattering, we have $\alpha \gg 1$. For
formal reasons, we also consider the case $\alpha \sim 1$ in this paper.

The expression for the width $w$ of the Lorentzian in Eq.~(\ref{dos1}) can 
be derived from a different point of view with the help of the golden
rule. With $|r>$ a resonance state and $|i>$ a state in the
$N$--dimensional subspace of Hilbert space comprising the GOE part of
${\bf H}$, we find, averaging over the states labelled $|i>$, 
\begin{equation}
w = \pi \frac{1}{N} \sum_i^N <r|V|i>^2 \delta (E_i-E_1) \ .
\label{the}
\end{equation}
Since $\sum_i^N \delta (E_i-E_1) = N \Delta / \pi \lambda$, we
immediately find $w = \Gamma\Delta$, i.e. the result we obtain using
the supersymmetry formalism. 

Using the explicit form of the matrix $Q^{*}$ in Eq.~(\ref{qmatrix}),
we can write the effective action in the exponent of Eq.~(\ref{simp2})
in the form 
\begin{equation}
-i\pi\frac{\omega}{d_{\rm eff}}{\rm trg}T^{-1}\Lambda T\Lambda ,
\label{simp1}
\end{equation}
where $d_{\rm eff}(E)$ is the inverse of the average level density 
$\overline{\rho (E)}$ given in Eq.~(\ref{dosob}). We note that, aside
from the definition of $d_{\rm eff}$, the ${\it form}$ of this
expression is the same as in the absence of any resonances. Taking
into account the fact that we can justify the second saddle-point
equation only for $\omega \ll \Gamma$, we conclude that (as in the
case of uncoupled resonances \cite{our}) the only change in the  
level-level correlation function caused by the presence of resonances
is due to the {\it rescaling of the mean level density}. Fig.~1 shows
the result for a set of parameters given in the figure caption.

\section{Conductance of a quasi one-dimensional sample}
\label{cond}

In this Section we consider a quasi one--dimensional system coupled to
two perfect leads. As mentioned in Sec.~\ref{M}, we consider this system
as consisting of $K$ slices labelled $i$ or $j$ with $i,j = 1, \ldots,
K$. Within each slice, the Hamiltonian has the form given in
Eq.~(2). Neighboring slices ($|j-i| = 1$) are coupled by random
matrices $H_{\mu\nu}^{ij}$ with Gaussian--distributed uncorrelated
matrix elements with zero mean value and a second moment given by
$(p^2/N) (\delta_{\mu, \mu'} \delta_{\nu, \nu'} + \delta_{\mu, \nu'}
\delta_{\nu, \mu'})$. For reasons given earlier, we assume that only
the random parts of the Hamiltonians in neighboring slices are coupled. 

In the case of a single slice, we have seen that the mutual coupling of 
resonances leads only to a quicker broadening of the resonances than would be 
the case otherwise, and finally to the disappearance of the resonance 
enhancement effect. With the intention of simplifying the presentation we,
therefore, now consider the situation where the resonances are not coupled 
even within one slice ($\lambda_1=0$). 

The coupling between the left (right) lead and the first (last) slice 
($j = 1$ or $K$) is described by the real matrix elements $W_{a\mu}^j$ 
\cite{Iid90}. Here, the index $a$ labels the channels in the leads. We 
assume that the dependence on energy of these matrix elements is negligible.
We are interested in the influence of the resonances on the coupling to 
the channels. Therefore, we allow for a direct coupling between leads 
and resonances. Hence, the index $\mu$ runs from $1$ to $N+m$. Following 
Ref. \cite{Iid90}, we assume for $j = 1,K$ the orthogonality relations
\begin{equation}
\sum_{\mu =1}^{N+m}W_{a\mu}^jW_{b\mu}^j = (N+m) \ (v_{a}^j)^{2} \ \delta_{ab}. 
\label{ortho}
\end{equation}

We use the many-channel Landauer formula
\begin{equation}
g=\sum_{a,b}(|S_{ab}^{LR}|^2+|S_{ab}^{RL}|^2) 
\label{gma1}
\end{equation}
which expresses the conductance $g$ at zero temperature in terms of the 
elements $S_{ab}^{LR}$ of the scattering matrix associated with the total 
Hamiltonian. Here $R$ and $L$ stand for the right and left leads, 
respectively. The scattering matrix has the form
\begin{equation}
S_{ab}^{cd}=\delta^{cd} \delta_{ab}-2i\pi\sum_{\mu
  ,\nu}W_{a\mu}^{c}[D^{-1}]_{\mu\nu}^{cd}W_{b\nu}^{d},
\label{sma}
\end{equation}
where the inverse propagator $D$ is a matrix of dimension
$K\times(N+m)$  given by
\begin{equation}
D_{\mu\nu}^{ij}=E\delta_{\mu\nu}\delta^{ij}-H_{\mu\nu}^{ij}+i\pi\sum_{a,c}
W_{a\mu}^{c}W_{a\nu}^{c}(\delta_{i1}\delta_{cL}+\delta_{iK}\delta_{cR}) 
\delta^{ij},
\label{dma}
\end{equation}
where $i, j=1,2,\ldots,K$ and $\mu, \nu = 1,2,\ldots, N+m$. The indices
$(c,d)$ on $D^{-1}$ in Eq.~(\ref{sma}) stand for (1,1) if $(c,d) = 
(L,L)$, for $(1,K)$ if $(c,d) = (L,R)$, etc. The energy $E$ equals the 
Fermi energy and eventually will be taken to have the value zero.

To calculate the ensemble average of $g$, we follow Refs.~\cite{Ver85} 
and \cite{Iid90} and express the conductance in terms of a supersymmetric 
generating functional. We only indicate the steps where we deviate from 
the references just given. We define a generating functional $Z$ for the 
conductance $g$, in a way similar to that presented in the previous section.  
\begin{equation}
Z(J)=\int {\cal D}[\Psi ] \exp\left[
\frac{1}{2}i
\Psi^{\dagger}L^{1/2}\left(E+iQ(j)
+i\eta L-{\bf H} +{\bf J}(j) \right)L^{1/2}\Psi\right] .
\label{gen2}
\end{equation}
The definitions of the source term ${\bf J}$ and of the matrix $Q$ may be 
found in Ref.~\cite{Iid90}. Suffice it to say that both ${\bf J}$ and $Q$ 
are different from zero only for the first and the last slice (only these 
two slices are coupled to the leads), and that $Q$ essentially comprises 
the last term on the rhs of Eq.~(\ref{dma}). After averaging and the 
Hubbard--Stratonovich transformation, the generating functional for the 
conductance $g$ has the form  
\begin{equation}
\overline{Z(J)} =
\int {\rm d} [\sigma]
\exp \left[-\frac{1}{4}\sum_{i,j}g_{ij}{\rm trg}\sigma (i)\sigma (j)-
\frac{1}{2}\sum_{j}{\rm Trg} \log 
(M(j)+iQ(j)+{\bf J}(j))\right] \ .
\label{fun}
\end{equation}
As before, the indices $i,j$ label the slices. The integration extends over 
$K$ independent $\sigma$--matrices. A matrix $g=P^{-1}$ of dimension $K$ 
(not to be confused with the conductance) is introduced, where 
$P_{ij}=(1/N)[\delta_{ij}+(p^2/\lambda^2)(\delta_{i,j+1}+\delta_{i+1,j})]$. 
For each $j$, the matrix $M(j)$ has the form given in Eq.~(\ref{mm}) with 
$\lambda_1 = 0$.

Using the saddle--point approximation (valid for $N\gg 1$), we determine 
the mean level density and show that the resonant structure found for a 
single slice remains qualitatively unchanged. Omitting the source term,
and taking the limit $m/N \rightarrow 0$, we obtain the saddle--point 
equation 
\begin{equation}
\sum_ig_{ij}\sigma (i)=\lambda \ N \frac{1}{E-\lambda \sigma (j)} \ .
\label{sd}
\end{equation}
We note that this equation is identical to the one which would obtain 
in the absence of all resonances. It was shown in Ref.~ \cite{Iid90} that 
the unique diagonal solution of Eq.~(\ref{sd}) has the form
\begin{equation}
\sigma_D(j)=r^j \ I_4-i\Delta^j \ L.
\label{solad}
\end{equation}
Assuming weak coupling between the slices, we find for the total density of 
states in the entire sample in zeroth order in $p^2/\lambda^2$
\begin{equation}
\overline{\rho}^{\rm tot} = \sum_{j=1}^{K}\overline{\rho}(j),
\label{dosdop}
\end{equation}
where $\overline{\rho}(j)$ is given by Eq.~(\ref{dos1}) with $r^j$ 
substituted for $E/2\lambda$, $\Delta^j$ for $\Delta$. This result shows 
that the Lorentzian part in the average level density which is due to the 
contribution from the resonances, is preserved. The importance of this 
statement lies in the fact that the conclusions valid for a single slice 
apply likewise to the entire sample. 

To see this, we calculate the two--point function. We follow Ref.~\cite{Iid90} 
and parameterize $\sigma (j)$ as $\sigma (j) = (T^j)^{-1}\sigma_D(j)T^j$ with  
\begin{equation}
T^j=\left(\begin{array}{cc}
\sqrt{1+t^{j}_{12}t^{j}_{21}} & it^{j}_{12} \\
-it^{j}_{21} & \sqrt{1+t^{j}_{21}t^{j}_{12}}\end{array}\right) \ .
\label{struc}
\end{equation}
We emphasize that in spite of the presence of the $m$ resonances in
each slice, the form of the matrices $\sigma(j)$ is ${\it the \ same}$
as in Ref.~\cite{Iid90} (absence of resonances). We recall that the
matrices $Q$ and ${\bf J}$ differ from zero for the first and the last
slice only. From the identity Eq.~(\ref{expo}) and the fact that ${\rm
Trg} \log \left (E-E_1 \right) = 0$, we conclude that the logarithmic
term in the exponent of Eq.~(\ref{fun}) vanishes unless $j = 1$ or $j
= K$. Except for these two terms, the resulting form of the generating
functional is identical to the one given in Ref.~\cite{Iid90}.
Moreover, the terms with $j = 1$ or $K$ differ from zero only because
of the presence of the matrices $Q$ and ${\bf J}$ both of which
involve the coupling to the channels. These results show that the
presence of resonances does not affect the behavior of the average
conductance in the bulk, but only the way it is coupled to the
leads. It was shown in Ref.~\cite{Iid90} that for sufficiently long
samples (where the Thouless energy is smaller than the decay width
defined by the coupling to the leads), the influence of the coupling
to the leads is immaterial for the behavior of the average
conductance. In practice, this condition is met for samples with
lengths exceeding a few elastic mean free paths. Using these facts, we
conclude that for typical samples, the average conductance is the same
with and without resonances. Since the average conductance is
calculated at ${\it fixed}$ frequency, this conclusion holds likewise
for electrons and for light (provided the frequency in the latter case
is sufficiently high). This is also true for the calculation of the
transmission coefficients given below. 

Before addressing the influence of the resonances on  the source term and on 
the coupling to the channels, we turn attention briefly to the intensity 
autocorrelation function ${\overline {g(E)g(E+\omega)}}$. We do not describe 
the calculation of this quantity (a four--point function) in any detail. It is 
not difficult to see, however, that the effective action for this
quantity also 
has the form given in Eq.~(\ref{fun}), except for the following changes. (i) 
The dimensions of the matrices $\sigma(j)$ are doubled. (ii) For each value of 
$j = 1, \ldots, K$, an extra term $- \omega \Lambda$ appears under the 
logarithm. Here, $\Lambda$ is the sixteen--dimensional analogue of the matrix 
defined above Eq.~(\ref{simp}). 

What are the changes caused by the appearance of this extra term under the 
logarithm? We neglect the source term and the $Q$ matrix. For each value of 
$j$, we expand the logarithm in powers of $\omega$, keeping only the linear 
term. (The term of zeroth order vanishes). After some simple algebra, we
find that the result has a form analogous to expression~(\ref{simp1}),
\begin{equation}
-i\pi\frac{\omega}{d_{\rm eff}} \sum_{j=1}^K {\rm trg}(T^j)^{-1} \Lambda T^j 
\Lambda \ .
\label{simp3}
\end{equation}
It has to be borne in mind that the $T$'s are now matrices of dimension 
sixteen. Here, $d_{\rm eff}$ is defined as the inverse of the mean level 
density ${\overline \rho}$ per slice, see Eq.~(\ref{dos1}), but with $r^j$ 
substituted for $E/2\lambda$, $\Delta^j$ for $\Delta$, see
Eq.~(\ref{solad}). We assume that ${\overline \rho}$ has the same
value in each slice. Comparing our result with what would apply in the
absence of resonances, we see that in each slice, $1 / d$ is replaced
by $1 / d_{\rm eff}$. For Efetov's effective action (obtained in the
continuum limit where the number of slices
tends to infinity while their individual length shrinks to zero) this means 
that the frequency $\omega$ is multiplied by the total level density 
${\overline \rho}^{\rm tot}$ defined in Eq.~(\ref{dosdop}). This is the claim 
made in Ref.~\cite{ourprl} and in the Introduction to the present paper. In 
contradistinction to the average conductance, the four--point correlation 
function ${\it is \ influenced}$ by the presence of resonances: The frequency 
$\omega$ occurs only in the combination $\omega {\overline \rho}^{\rm tot}$. 
The result for ${\cal D}$ is shown in Figure~1 of Ref.~\cite{ourprl}. 
The argument can easily be extended to correlation functions of higher order.

The conclusions drawn in the last three paragraphs contain the main results 
of this work. It remains to calculate the influence of the resonances on the 
coupling to the leads. For reasons given above, this may be considered a 
somewhat academic exercise. However, experiments on short samples would be 
able to test the predictions made in this way.

The terms describing the coupling to the channels are part 
of the matrices $Q$ and ${\bf J}$ under the logarithm in Eq.~(\ref{fun}). We 
recall that both these terms are different from zero only for the first ($j 
= 1$) and last ($j = K$) slices. In expanding the logarithm, we have to keep 
terms of any order in $iQ(j)$ but only of zeroth and first order in ${\bf J}$. 
Here, we focus attention on the terms of zeroth order in ${\bf J}$. (The 
calculation of the first--order terms proceeds analogously). After calculating 
the trace over $\mu$ in each term of the series by using the orthogonality 
relation Eq.~(\ref{ortho}) for $j=1$ ($K$) and resumming the result, we use
the saddle--point solution for the first and the last slices, the 
parameterization of $\sigma$ given in Eq.~(\ref{struc}), and introduce the 
quantities $\Gamma_{a} = V_{a}^2 / \lambda$. After some algebra, we
find for $j = 1,K$ 
\begin{equation}
-\frac{1}{2}\sum_a{\rm Trg}
\log\left(1+\frac{1}{2}T_at^j_{12}t^j_{21}\right),
\label{sti1}
\end{equation}
where $T_a$ is the "sticking probability" in channel $a$, given by
\begin{equation}
T_a=\frac{4\Delta\left(-\frac{\Gamma^2_a}{\lambda}\frac{\Gamma-\frac{E}
{\lambda}(E-E_1)}{(E-E_1-\frac{E}{2\lambda}\Gamma)^2+\Gamma^2\Delta^2}+
\frac{W_a}{\lambda}\right)}
{\left(1+\frac{W_a}{\lambda}\Delta-\frac{\Gamma^2_a}
{\lambda}\Delta\frac{\Gamma-\frac{E}{\lambda}(E-E_1)}{(E-E_1-\frac{E} 
{2\lambda}\Gamma)^2+\Gamma^2\Delta^2}\right)^2+
\left(\frac{E}{2\lambda}\frac{W_a}{\lambda}
-\frac{\Gamma^2_a}{\lambda}\frac{(E-E_1)(1-\frac{1}{2}(\frac{E}{\lambda})^2
+\frac{E}{2\lambda}\Gamma}{(E-E_1-\frac{E}{2\lambda}\Gamma)^2+\Gamma^2
\Delta^2}\right)^2} \ .
\label{sti2}
\end{equation}
The sticking probabilities $T_a$ are related to the matrix elements 
$\overline{S_{aa}^{cc}}$ of the average scattering matrix by $T_a=1-| 
\overline{S_{aa}^{cc}}|^2$. This means that 
$T_a$ measures that part of the flux which is not re-emitted instantaneously 
into the same channel but which experiences multiple scattering in the 
disordered sample. Eq.~(\ref{sti2}) very clearly shows that the sticking 
probabilities are affected by a direct coupling between resonances and 
channels. 

\section{Conclusions}
\label{concl}

It was shown in Ref. \cite{Iid90} that the conductance of a short 
sample is proportional to the product of two terms, each term being
given by the sum of the sticking probabilities of the channels in one
of the two leads. Thus, the conductance of a short sample is
determined by the coupling to the leads, whereas the conductance of a
long sample is entirely determined by the matrix $g_{ij}$ appearing
in Eq.~(\ref{fun}). This implies that as long as there is no direct
coupling between resonances and leads, the conductances of both short
and long samples are unchanged. When the resonances are coupled to the
channels, however, the conductance of a short sample can be changed
drastically at frequencies near the resonance frequency, whereas the
corresponding change of the conductance of a long sample is expected
to be small. It would be interesting to search for this effect
experimentally. 
 
On the other hand, the intensity autocorrelation function is {\it always}
modified by the presence of resonances, both for short and for long
samples. This fact follows directly from our calculation of the
diffusion constant ${\cal D}$ (which shows a resonance dip) but is most
easily seen by using an argument based on the Thouless relation. This
argument was given in the Introduction and is not repeated here. The
argument is fully corroborated by our analytical results.

It was emphasized above that the effective Lagrangean for electrons
and for scalar waves (light) is the same, both in the absence and in
the presence of Mie resonances, provided that the usual condition $k l
\gg 1$ is met. Here, $k$ is the wave number and $l$ the elastic mean
free path. In view of remarks in the literature that the strong effect
of Mie resonances upon light scattering is absent in the case of
electrons, we wish to clarify this statement. In the case of scalar
waves (light), there are two different and unrelated effects which
contribute towards the energy dependence of the potential. The first
of these is due to the dependence on $k$ of the term $k^2
\epsilon(\vec r)$ in the scalar wave equation.  Here, $\epsilon$ is
the space--dependent dielectric constant. The disorder potential $V$
in the Schr\"odinger equation has no analogous dependence on wave
number or energy. However, the change with $k$ of $k^2 \epsilon(\vec
r)$ is small on the scale of the mean level spacing. This is why the
non-linear sigma models for electrons and for scalar waves are
identical, even though the Ward identities for the two wave equations
differ. Second, the presence of Mie resonances causes the drastic
modification of the intensity autocorrelation function for light
scattering discussed above. The equality of the nonlinear sigma models
for scalar and for Schr\"odinger waves implies that a correspondingly
large modification of the intensity autocorrelation function would
also occur for electrons {\it if it were possible to design a system
  in which electrons are scattered both diffusively and via Mie--like
  resonances}. To the best of our knowledge, such a system does not
exist at present, and the stated analogy makes a prediction which
cannot be checked experimentally.

To compare with experiment, we note: (i) For classical waves in a uniform
medium in $d$ dimensions, the density of states is proportional to
$\epsilon^{d/2}$ where $\epsilon$ is the dielectric constant of the medium. 
Increasing the number of resonating spheres, we increase the effective
dielectric constant of the medium. Hence, $\lambda$ should decrease with
increasing $\mu$. (ii) As shown in Sec.~\ref{M}, $\lambda_1$ is
expected to increase monotonically with $\mu$. Both these points
affect the dependence on concentration of $d_{\rm eff}$ and,
therefore, of ${\cal D} \sim d_{\rm eff}$. With increasing $\mu$,
point (i) causes a decrease of ${\cal D}$ far away from the resonance
dip, and point (ii) a widening the resonance dip. Fig.~1 of
Ref.~\cite{ourprl} shows ${\cal D}$ versus energy for different
concentrations. In agreement with the experimental results
\cite{Alb91,Gar}, we find a deep dip in the diffusion constant at the
resonance energy. It is caused by the second term in Eq.~(\ref{dos1})
and is particularly pronounced at low concentrations of the resonant
scatterers. As the concentration increases, the dip is smeared out. At
the same time, the value of ${\cal D}$ outside the resonance dip
decreases.
 
\section{Acknowledgments}Discussions with Dr. A. M\"uller-Groeling
led to the central idea of how to treat the resonances in the
framework of random matrix theory. V. K. appreciates very useful
discussions with Dr. Y. Fyodorov and gratefully acknowledges the
support of a MINERVA Fellowship. B. E. wishes to thank
Prof. A. Nourreddine for valuable discussions.

\begin{figure}
\label{fig1}
\input{greatfig1}
\vspace*{1cm}
\caption{The exact normalized level--level correlation function 
for $m=5$, $\Gamma =5d$, $E=E_1=0$ and rescaled correlation 
function for a pure GUE given by
$\sin^2(2\pi\omega /d_{eff})/(2\pi\omega /d_{eff})^2$.}

\end{figure}
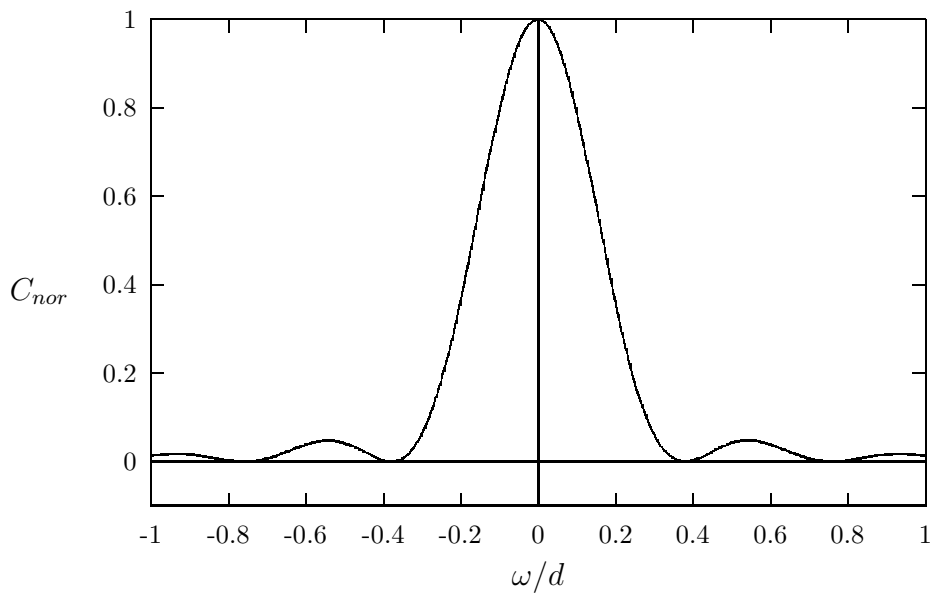

\end{document}

%% file: greatfig1.tex
\setlength{\unitlength}{0.240900pt}
\ifx\plotpoint\undefined\newsavebox{\plotpoint}\fi
\sbox{\plotpoint}{\rule[-0.200pt]{0.400pt}{0.400pt}}%
\begin{picture}(1500,900)(0,0)
\font\gnuplot=cmr10 at 10pt
\gnuplot
\sbox{\plotpoint}{\rule[-0.200pt]{0.400pt}{0.400pt}}%
\put(220.0,182.0){\rule[-0.200pt]{292.934pt}{0.400pt}}
\put(828.0,113.0){\rule[-0.200pt]{0.400pt}{184.048pt}}
\put(220.0,182.0){\rule[-0.200pt]{4.818pt}{0.400pt}}
\put(198,182){\makebox(0,0)[r]{0}}
\put(1416.0,182.0){\rule[-0.200pt]{4.818pt}{0.400pt}}
\put(220.0,321.0){\rule[-0.200pt]{4.818pt}{0.400pt}}
\put(198,321){\makebox(0,0)[r]{0.2}}
\put(1416.0,321.0){\rule[-0.200pt]{4.818pt}{0.400pt}}
\put(220.0,460.0){\rule[-0.200pt]{4.818pt}{0.400pt}}
\put(198,460){\makebox(0,0)[r]{0.4}}
\put(1416.0,460.0){\rule[-0.200pt]{4.818pt}{0.400pt}}
\put(220.0,599.0){\rule[-0.200pt]{4.818pt}{0.400pt}}
\put(198,599){\makebox(0,0)[r]{0.6}}
\put(1416.0,599.0){\rule[-0.200pt]{4.818pt}{0.400pt}}
\put(220.0,738.0){\rule[-0.200pt]{4.818pt}{0.400pt}}
\put(198,738){\makebox(0,0)[r]{0.8}}
\put(1416.0,738.0){\rule[-0.200pt]{4.818pt}{0.400pt}}
\put(220.0,877.0){\rule[-0.200pt]{4.818pt}{0.400pt}}
\put(198,877){\makebox(0,0)[r]{1}}
\put(1416.0,877.0){\rule[-0.200pt]{4.818pt}{0.400pt}}
\put(220.0,113.0){\rule[-0.200pt]{0.400pt}{4.818pt}}
\put(220,68){\makebox(0,0){-1}}
\put(220.0,857.0){\rule[-0.200pt]{0.400pt}{4.818pt}}
\put(342.0,113.0){\rule[-0.200pt]{0.400pt}{4.818pt}}
\put(342,68){\makebox(0,0){-0.8}}
\put(342.0,857.0){\rule[-0.200pt]{0.400pt}{4.818pt}}
\put(463.0,113.0){\rule[-0.200pt]{0.400pt}{4.818pt}}
\put(463,68){\makebox(0,0){-0.6}}
\put(463.0,857.0){\rule[-0.200pt]{0.400pt}{4.818pt}}
\put(585.0,113.0){\rule[-0.200pt]{0.400pt}{4.818pt}}
\put(585,68){\makebox(0,0){-0.4}}
\put(585.0,857.0){\rule[-0.200pt]{0.400pt}{4.818pt}}
\put(706.0,113.0){\rule[-0.200pt]{0.400pt}{4.818pt}}
\put(706,68){\makebox(0,0){-0.2}}
\put(706.0,857.0){\rule[-0.200pt]{0.400pt}{4.818pt}}
\put(828.0,113.0){\rule[-0.200pt]{0.400pt}{4.818pt}}
\put(828,68){\makebox(0,0){0}}
\put(828.0,857.0){\rule[-0.200pt]{0.400pt}{4.818pt}}
\put(950.0,113.0){\rule[-0.200pt]{0.400pt}{4.818pt}}
\put(950,68){\makebox(0,0){0.2}}
\put(950.0,857.0){\rule[-0.200pt]{0.400pt}{4.818pt}}
\put(1071.0,113.0){\rule[-0.200pt]{0.400pt}{4.818pt}}
\put(1071,68){\makebox(0,0){0.4}}
\put(1071.0,857.0){\rule[-0.200pt]{0.400pt}{4.818pt}}
\put(1193.0,113.0){\rule[-0.200pt]{0.400pt}{4.818pt}}
\put(1193,68){\makebox(0,0){0.6}}
\put(1193.0,857.0){\rule[-0.200pt]{0.400pt}{4.818pt}}
\put(1314.0,113.0){\rule[-0.200pt]{0.400pt}{4.818pt}}
\put(1314,68){\makebox(0,0){0.8}}
\put(1314.0,857.0){\rule[-0.200pt]{0.400pt}{4.818pt}}
\put(1436.0,113.0){\rule[-0.200pt]{0.400pt}{4.818pt}}
\put(1436,68){\makebox(0,0){1}}
\put(1436.0,857.0){\rule[-0.200pt]{0.400pt}{4.818pt}}
\put(220.0,113.0){\rule[-0.200pt]{292.934pt}{0.400pt}}
\put(1436.0,113.0){\rule[-0.200pt]{0.400pt}{184.048pt}}
\put(220.0,877.0){\rule[-0.200pt]{292.934pt}{0.400pt}}
\put(45,450){\makebox(0,0){$C_{nor}$}}
\put(828,0){\makebox(0,0){$\omega /d$}}
\put(220.0,113.0){\rule[-0.200pt]{0.400pt}{184.048pt}}
\put(220,190.67){\rule{1.204pt}{0.400pt}}
\multiput(220.00,190.17)(2.500,1.000){2}{\rule{0.602pt}{0.400pt}}
\put(232,191.67){\rule{1.445pt}{0.400pt}}
\multiput(232.00,191.17)(3.000,1.000){2}{\rule{0.723pt}{0.400pt}}
\put(225.0,192.0){\rule[-0.200pt]{1.686pt}{0.400pt}}
\put(244,192.67){\rule{1.445pt}{0.400pt}}
\multiput(244.00,192.17)(3.000,1.000){2}{\rule{0.723pt}{0.400pt}}
\put(238.0,193.0){\rule[-0.200pt]{1.445pt}{0.400pt}}
\put(274,192.67){\rule{1.445pt}{0.400pt}}
\multiput(274.00,193.17)(3.000,-1.000){2}{\rule{0.723pt}{0.400pt}}
\put(250.0,194.0){\rule[-0.200pt]{5.782pt}{0.400pt}}
\put(286,191.67){\rule{1.445pt}{0.400pt}}
\multiput(286.00,192.17)(3.000,-1.000){2}{\rule{0.723pt}{0.400pt}}
\put(292,190.67){\rule{1.445pt}{0.400pt}}
\multiput(292.00,191.17)(3.000,-1.000){2}{\rule{0.723pt}{0.400pt}}
\put(298,189.67){\rule{1.686pt}{0.400pt}}
\multiput(298.00,190.17)(3.500,-1.000){2}{\rule{0.843pt}{0.400pt}}
\put(305,188.67){\rule{1.445pt}{0.400pt}}
\multiput(305.00,189.17)(3.000,-1.000){2}{\rule{0.723pt}{0.400pt}}
\put(311,187.67){\rule{1.445pt}{0.400pt}}
\multiput(311.00,188.17)(3.000,-1.000){2}{\rule{0.723pt}{0.400pt}}
\put(317,186.67){\rule{1.445pt}{0.400pt}}
\multiput(317.00,187.17)(3.000,-1.000){2}{\rule{0.723pt}{0.400pt}}
\put(323,185.67){\rule{1.445pt}{0.400pt}}
\multiput(323.00,186.17)(3.000,-1.000){2}{\rule{0.723pt}{0.400pt}}
\put(329,184.67){\rule{1.445pt}{0.400pt}}
\multiput(329.00,185.17)(3.000,-1.000){2}{\rule{0.723pt}{0.400pt}}
\put(335,183.67){\rule{1.445pt}{0.400pt}}
\multiput(335.00,184.17)(3.000,-1.000){2}{\rule{0.723pt}{0.400pt}}
\put(280.0,193.0){\rule[-0.200pt]{1.445pt}{0.400pt}}
\put(347,182.67){\rule{1.445pt}{0.400pt}}
\multiput(347.00,183.17)(3.000,-1.000){2}{\rule{0.723pt}{0.400pt}}
\put(341.0,184.0){\rule[-0.200pt]{1.445pt}{0.400pt}}
\put(359,181.67){\rule{1.445pt}{0.400pt}}
\multiput(359.00,182.17)(3.000,-1.000){2}{\rule{0.723pt}{0.400pt}}
\put(365,181.67){\rule{1.445pt}{0.400pt}}
\multiput(365.00,181.17)(3.000,1.000){2}{\rule{0.723pt}{0.400pt}}
\put(353.0,183.0){\rule[-0.200pt]{1.445pt}{0.400pt}}
\put(384,182.67){\rule{1.445pt}{0.400pt}}
\multiput(384.00,182.17)(3.000,1.000){2}{\rule{0.723pt}{0.400pt}}
\put(390,183.67){\rule{1.445pt}{0.400pt}}
\multiput(390.00,183.17)(3.000,1.000){2}{\rule{0.723pt}{0.400pt}}
\put(396,185.17){\rule{1.300pt}{0.400pt}}
\multiput(396.00,184.17)(3.302,2.000){2}{\rule{0.650pt}{0.400pt}}
\put(402,186.67){\rule{1.445pt}{0.400pt}}
\multiput(402.00,186.17)(3.000,1.000){2}{\rule{0.723pt}{0.400pt}}
\put(408,188.17){\rule{1.300pt}{0.400pt}}
\multiput(408.00,187.17)(3.302,2.000){2}{\rule{0.650pt}{0.400pt}}
\put(414,190.17){\rule{1.300pt}{0.400pt}}
\multiput(414.00,189.17)(3.302,2.000){2}{\rule{0.650pt}{0.400pt}}
\multiput(420.00,192.61)(1.132,0.447){3}{\rule{0.900pt}{0.108pt}}
\multiput(420.00,191.17)(4.132,3.000){2}{\rule{0.450pt}{0.400pt}}
\put(426,195.17){\rule{1.300pt}{0.400pt}}
\multiput(426.00,194.17)(3.302,2.000){2}{\rule{0.650pt}{0.400pt}}
\put(432,197.17){\rule{1.300pt}{0.400pt}}
\multiput(432.00,196.17)(3.302,2.000){2}{\rule{0.650pt}{0.400pt}}
\multiput(438.00,199.61)(1.132,0.447){3}{\rule{0.900pt}{0.108pt}}
\multiput(438.00,198.17)(4.132,3.000){2}{\rule{0.450pt}{0.400pt}}
\put(444,202.17){\rule{1.300pt}{0.400pt}}
\multiput(444.00,201.17)(3.302,2.000){2}{\rule{0.650pt}{0.400pt}}
\put(450,204.17){\rule{1.500pt}{0.400pt}}
\multiput(450.00,203.17)(3.887,2.000){2}{\rule{0.750pt}{0.400pt}}
\multiput(457.00,206.61)(1.132,0.447){3}{\rule{0.900pt}{0.108pt}}
\multiput(457.00,205.17)(4.132,3.000){2}{\rule{0.450pt}{0.400pt}}
\put(463,208.67){\rule{1.445pt}{0.400pt}}
\multiput(463.00,208.17)(3.000,1.000){2}{\rule{0.723pt}{0.400pt}}
\put(469,210.17){\rule{1.300pt}{0.400pt}}
\multiput(469.00,209.17)(3.302,2.000){2}{\rule{0.650pt}{0.400pt}}
\put(475,211.67){\rule{1.445pt}{0.400pt}}
\multiput(475.00,211.17)(3.000,1.000){2}{\rule{0.723pt}{0.400pt}}
\put(481,212.67){\rule{1.445pt}{0.400pt}}
\multiput(481.00,212.17)(3.000,1.000){2}{\rule{0.723pt}{0.400pt}}
\put(487,213.67){\rule{1.445pt}{0.400pt}}
\multiput(487.00,213.17)(3.000,1.000){2}{\rule{0.723pt}{0.400pt}}
\put(371.0,183.0){\rule[-0.200pt]{3.132pt}{0.400pt}}
\put(505,213.67){\rule{1.445pt}{0.400pt}}
\multiput(505.00,214.17)(3.000,-1.000){2}{\rule{0.723pt}{0.400pt}}
\put(511,212.67){\rule{1.445pt}{0.400pt}}
\multiput(511.00,213.17)(3.000,-1.000){2}{\rule{0.723pt}{0.400pt}}
\put(517,211.17){\rule{1.300pt}{0.400pt}}
\multiput(517.00,212.17)(3.302,-2.000){2}{\rule{0.650pt}{0.400pt}}
\put(523,209.17){\rule{1.300pt}{0.400pt}}
\multiput(523.00,210.17)(3.302,-2.000){2}{\rule{0.650pt}{0.400pt}}
\put(529,207.17){\rule{1.500pt}{0.400pt}}
\multiput(529.00,208.17)(3.887,-2.000){2}{\rule{0.750pt}{0.400pt}}
\multiput(536.00,205.95)(1.132,-0.447){3}{\rule{0.900pt}{0.108pt}}
\multiput(536.00,206.17)(4.132,-3.000){2}{\rule{0.450pt}{0.400pt}}
\multiput(542.00,202.95)(1.132,-0.447){3}{\rule{0.900pt}{0.108pt}}
\multiput(542.00,203.17)(4.132,-3.000){2}{\rule{0.450pt}{0.400pt}}
\multiput(548.00,199.95)(1.132,-0.447){3}{\rule{0.900pt}{0.108pt}}
\multiput(548.00,200.17)(4.132,-3.000){2}{\rule{0.450pt}{0.400pt}}
\multiput(554.00,196.95)(1.132,-0.447){3}{\rule{0.900pt}{0.108pt}}
\multiput(554.00,197.17)(4.132,-3.000){2}{\rule{0.450pt}{0.400pt}}
\multiput(560.00,193.95)(1.132,-0.447){3}{\rule{0.900pt}{0.108pt}}
\multiput(560.00,194.17)(4.132,-3.000){2}{\rule{0.450pt}{0.400pt}}
\multiput(566.00,190.95)(1.132,-0.447){3}{\rule{0.900pt}{0.108pt}}
\multiput(566.00,191.17)(4.132,-3.000){2}{\rule{0.450pt}{0.400pt}}
\multiput(572.00,187.95)(1.132,-0.447){3}{\rule{0.900pt}{0.108pt}}
\multiput(572.00,188.17)(4.132,-3.000){2}{\rule{0.450pt}{0.400pt}}
\put(578,184.17){\rule{1.300pt}{0.400pt}}
\multiput(578.00,185.17)(3.302,-2.000){2}{\rule{0.650pt}{0.400pt}}
\put(584,182.67){\rule{1.445pt}{0.400pt}}
\multiput(584.00,183.17)(3.000,-1.000){2}{\rule{0.723pt}{0.400pt}}
\put(590,181.67){\rule{1.445pt}{0.400pt}}
\multiput(590.00,182.17)(3.000,-1.000){2}{\rule{0.723pt}{0.400pt}}
\put(596,181.67){\rule{1.445pt}{0.400pt}}
\multiput(596.00,181.17)(3.000,1.000){2}{\rule{0.723pt}{0.400pt}}
\put(602,182.67){\rule{1.686pt}{0.400pt}}
\multiput(602.00,182.17)(3.500,1.000){2}{\rule{0.843pt}{0.400pt}}
\multiput(609.00,184.61)(1.132,0.447){3}{\rule{0.900pt}{0.108pt}}
\multiput(609.00,183.17)(4.132,3.000){2}{\rule{0.450pt}{0.400pt}}
\multiput(615.00,187.60)(0.774,0.468){5}{\rule{0.700pt}{0.113pt}}
\multiput(615.00,186.17)(4.547,4.000){2}{\rule{0.350pt}{0.400pt}}
\multiput(621.00,191.59)(0.599,0.477){7}{\rule{0.580pt}{0.115pt}}
\multiput(621.00,190.17)(4.796,5.000){2}{\rule{0.290pt}{0.400pt}}
\multiput(627.59,196.00)(0.482,0.671){9}{\rule{0.116pt}{0.633pt}}
\multiput(626.17,196.00)(6.000,6.685){2}{\rule{0.400pt}{0.317pt}}
\multiput(633.59,204.00)(0.482,0.762){9}{\rule{0.116pt}{0.700pt}}
\multiput(632.17,204.00)(6.000,7.547){2}{\rule{0.400pt}{0.350pt}}
\multiput(639.59,213.00)(0.482,0.852){9}{\rule{0.116pt}{0.767pt}}
\multiput(638.17,213.00)(6.000,8.409){2}{\rule{0.400pt}{0.383pt}}
\multiput(645.59,223.00)(0.482,1.123){9}{\rule{0.116pt}{0.967pt}}
\multiput(644.17,223.00)(6.000,10.994){2}{\rule{0.400pt}{0.483pt}}
\multiput(651.59,236.00)(0.482,1.214){9}{\rule{0.116pt}{1.033pt}}
\multiput(650.17,236.00)(6.000,11.855){2}{\rule{0.400pt}{0.517pt}}
\multiput(657.59,250.00)(0.482,1.485){9}{\rule{0.116pt}{1.233pt}}
\multiput(656.17,250.00)(6.000,14.440){2}{\rule{0.400pt}{0.617pt}}
\multiput(663.59,267.00)(0.482,1.575){9}{\rule{0.116pt}{1.300pt}}
\multiput(662.17,267.00)(6.000,15.302){2}{\rule{0.400pt}{0.650pt}}
\multiput(669.59,285.00)(0.482,1.756){9}{\rule{0.116pt}{1.433pt}}
\multiput(668.17,285.00)(6.000,17.025){2}{\rule{0.400pt}{0.717pt}}
\multiput(675.59,305.00)(0.482,1.937){9}{\rule{0.116pt}{1.567pt}}
\multiput(674.17,305.00)(6.000,18.748){2}{\rule{0.400pt}{0.783pt}}
\multiput(681.59,327.00)(0.485,1.789){11}{\rule{0.117pt}{1.471pt}}
\multiput(680.17,327.00)(7.000,20.946){2}{\rule{0.400pt}{0.736pt}}
\multiput(688.59,351.00)(0.482,2.208){9}{\rule{0.116pt}{1.767pt}}
\multiput(687.17,351.00)(6.000,21.333){2}{\rule{0.400pt}{0.883pt}}
\multiput(694.59,376.00)(0.482,2.389){9}{\rule{0.116pt}{1.900pt}}
\multiput(693.17,376.00)(6.000,23.056){2}{\rule{0.400pt}{0.950pt}}
\multiput(700.59,403.00)(0.482,2.480){9}{\rule{0.116pt}{1.967pt}}
\multiput(699.17,403.00)(6.000,23.918){2}{\rule{0.400pt}{0.983pt}}
\multiput(706.59,431.00)(0.482,2.570){9}{\rule{0.116pt}{2.033pt}}
\multiput(705.17,431.00)(6.000,24.780){2}{\rule{0.400pt}{1.017pt}}
\multiput(712.59,460.00)(0.482,2.660){9}{\rule{0.116pt}{2.100pt}}
\multiput(711.17,460.00)(6.000,25.641){2}{\rule{0.400pt}{1.050pt}}
\multiput(718.59,490.00)(0.482,2.660){9}{\rule{0.116pt}{2.100pt}}
\multiput(717.17,490.00)(6.000,25.641){2}{\rule{0.400pt}{1.050pt}}
\multiput(724.59,520.00)(0.482,2.751){9}{\rule{0.116pt}{2.167pt}}
\multiput(723.17,520.00)(6.000,26.503){2}{\rule{0.400pt}{1.083pt}}
\multiput(730.59,551.00)(0.482,2.751){9}{\rule{0.116pt}{2.167pt}}
\multiput(729.17,551.00)(6.000,26.503){2}{\rule{0.400pt}{1.083pt}}
\multiput(736.59,582.00)(0.482,2.751){9}{\rule{0.116pt}{2.167pt}}
\multiput(735.17,582.00)(6.000,26.503){2}{\rule{0.400pt}{1.083pt}}
\multiput(742.59,613.00)(0.482,2.751){9}{\rule{0.116pt}{2.167pt}}
\multiput(741.17,613.00)(6.000,26.503){2}{\rule{0.400pt}{1.083pt}}
\multiput(748.59,644.00)(0.482,2.570){9}{\rule{0.116pt}{2.033pt}}
\multiput(747.17,644.00)(6.000,24.780){2}{\rule{0.400pt}{1.017pt}}
\multiput(754.59,673.00)(0.485,2.171){11}{\rule{0.117pt}{1.757pt}}
\multiput(753.17,673.00)(7.000,25.353){2}{\rule{0.400pt}{0.879pt}}
\multiput(761.59,702.00)(0.482,2.389){9}{\rule{0.116pt}{1.900pt}}
\multiput(760.17,702.00)(6.000,23.056){2}{\rule{0.400pt}{0.950pt}}
\multiput(767.59,729.00)(0.482,2.299){9}{\rule{0.116pt}{1.833pt}}
\multiput(766.17,729.00)(6.000,22.195){2}{\rule{0.400pt}{0.917pt}}
\multiput(773.59,755.00)(0.482,2.118){9}{\rule{0.116pt}{1.700pt}}
\multiput(772.17,755.00)(6.000,20.472){2}{\rule{0.400pt}{0.850pt}}
\multiput(779.59,779.00)(0.482,1.937){9}{\rule{0.116pt}{1.567pt}}
\multiput(778.17,779.00)(6.000,18.748){2}{\rule{0.400pt}{0.783pt}}
\multiput(785.59,801.00)(0.482,1.666){9}{\rule{0.116pt}{1.367pt}}
\multiput(784.17,801.00)(6.000,16.163){2}{\rule{0.400pt}{0.683pt}}
\multiput(791.59,820.00)(0.482,1.485){9}{\rule{0.116pt}{1.233pt}}
\multiput(790.17,820.00)(6.000,14.440){2}{\rule{0.400pt}{0.617pt}}
\multiput(797.59,837.00)(0.482,1.214){9}{\rule{0.116pt}{1.033pt}}
\multiput(796.17,837.00)(6.000,11.855){2}{\rule{0.400pt}{0.517pt}}
\multiput(803.59,851.00)(0.482,0.943){9}{\rule{0.116pt}{0.833pt}}
\multiput(802.17,851.00)(6.000,9.270){2}{\rule{0.400pt}{0.417pt}}
\multiput(809.59,862.00)(0.482,0.671){9}{\rule{0.116pt}{0.633pt}}
\multiput(808.17,862.00)(6.000,6.685){2}{\rule{0.400pt}{0.317pt}}
\multiput(815.00,870.59)(0.599,0.477){7}{\rule{0.580pt}{0.115pt}}
\multiput(815.00,869.17)(4.796,5.000){2}{\rule{0.290pt}{0.400pt}}
\put(821,875.17){\rule{1.300pt}{0.400pt}}
\multiput(821.00,874.17)(3.302,2.000){2}{\rule{0.650pt}{0.400pt}}
\put(827,875.67){\rule{1.445pt}{0.400pt}}
\multiput(827.00,876.17)(3.000,-1.000){2}{\rule{0.723pt}{0.400pt}}
\multiput(833.00,874.93)(0.710,-0.477){7}{\rule{0.660pt}{0.115pt}}
\multiput(833.00,875.17)(5.630,-5.000){2}{\rule{0.330pt}{0.400pt}}
\multiput(840.59,868.65)(0.482,-0.581){9}{\rule{0.116pt}{0.567pt}}
\multiput(839.17,869.82)(6.000,-5.824){2}{\rule{0.400pt}{0.283pt}}
\multiput(846.59,860.54)(0.482,-0.943){9}{\rule{0.116pt}{0.833pt}}
\multiput(845.17,862.27)(6.000,-9.270){2}{\rule{0.400pt}{0.417pt}}
\multiput(852.59,848.99)(0.482,-1.123){9}{\rule{0.116pt}{0.967pt}}
\multiput(851.17,850.99)(6.000,-10.994){2}{\rule{0.400pt}{0.483pt}}
\multiput(858.59,834.88)(0.482,-1.485){9}{\rule{0.116pt}{1.233pt}}
\multiput(857.17,837.44)(6.000,-14.440){2}{\rule{0.400pt}{0.617pt}}
\multiput(864.59,817.60)(0.482,-1.575){9}{\rule{0.116pt}{1.300pt}}
\multiput(863.17,820.30)(6.000,-15.302){2}{\rule{0.400pt}{0.650pt}}
\multiput(870.59,798.50)(0.482,-1.937){9}{\rule{0.116pt}{1.567pt}}
\multiput(869.17,801.75)(6.000,-18.748){2}{\rule{0.400pt}{0.783pt}}
\multiput(876.59,776.22)(0.482,-2.027){9}{\rule{0.116pt}{1.633pt}}
\multiput(875.17,779.61)(6.000,-19.610){2}{\rule{0.400pt}{0.817pt}}
\multiput(882.59,752.67)(0.482,-2.208){9}{\rule{0.116pt}{1.767pt}}
\multiput(881.17,756.33)(6.000,-21.333){2}{\rule{0.400pt}{0.883pt}}
\multiput(888.59,727.11)(0.482,-2.389){9}{\rule{0.116pt}{1.900pt}}
\multiput(887.17,731.06)(6.000,-23.056){2}{\rule{0.400pt}{0.950pt}}
\multiput(894.59,699.56)(0.482,-2.570){9}{\rule{0.116pt}{2.033pt}}
\multiput(893.17,703.78)(6.000,-24.780){2}{\rule{0.400pt}{1.017pt}}
\multiput(900.59,670.56)(0.482,-2.570){9}{\rule{0.116pt}{2.033pt}}
\multiput(899.17,674.78)(6.000,-24.780){2}{\rule{0.400pt}{1.017pt}}
\multiput(906.59,642.23)(0.485,-2.323){11}{\rule{0.117pt}{1.871pt}}
\multiput(905.17,646.12)(7.000,-27.116){2}{\rule{0.400pt}{0.936pt}}
\multiput(913.59,610.28)(0.482,-2.660){9}{\rule{0.116pt}{2.100pt}}
\multiput(912.17,614.64)(6.000,-25.641){2}{\rule{0.400pt}{1.050pt}}
\multiput(919.59,580.01)(0.482,-2.751){9}{\rule{0.116pt}{2.167pt}}
\multiput(918.17,584.50)(6.000,-26.503){2}{\rule{0.400pt}{1.083pt}}
\multiput(925.59,549.01)(0.482,-2.751){9}{\rule{0.116pt}{2.167pt}}
\multiput(924.17,553.50)(6.000,-26.503){2}{\rule{0.400pt}{1.083pt}}
\multiput(931.59,518.01)(0.482,-2.751){9}{\rule{0.116pt}{2.167pt}}
\multiput(930.17,522.50)(6.000,-26.503){2}{\rule{0.400pt}{1.083pt}}
\multiput(937.59,487.28)(0.482,-2.660){9}{\rule{0.116pt}{2.100pt}}
\multiput(936.17,491.64)(6.000,-25.641){2}{\rule{0.400pt}{1.050pt}}
\multiput(943.59,457.56)(0.482,-2.570){9}{\rule{0.116pt}{2.033pt}}
\multiput(942.17,461.78)(6.000,-24.780){2}{\rule{0.400pt}{1.017pt}}
\multiput(949.59,428.56)(0.482,-2.570){9}{\rule{0.116pt}{2.033pt}}
\multiput(948.17,432.78)(6.000,-24.780){2}{\rule{0.400pt}{1.017pt}}
\multiput(955.59,400.11)(0.482,-2.389){9}{\rule{0.116pt}{1.900pt}}
\multiput(954.17,404.06)(6.000,-23.056){2}{\rule{0.400pt}{0.950pt}}
\multiput(961.59,373.67)(0.482,-2.208){9}{\rule{0.116pt}{1.767pt}}
\multiput(960.17,377.33)(6.000,-21.333){2}{\rule{0.400pt}{0.883pt}}
\multiput(967.59,348.94)(0.482,-2.118){9}{\rule{0.116pt}{1.700pt}}
\multiput(966.17,352.47)(6.000,-20.472){2}{\rule{0.400pt}{0.850pt}}
\multiput(973.59,325.22)(0.482,-2.027){9}{\rule{0.116pt}{1.633pt}}
\multiput(972.17,328.61)(6.000,-19.610){2}{\rule{0.400pt}{0.817pt}}
\multiput(979.59,303.05)(0.482,-1.756){9}{\rule{0.116pt}{1.433pt}}
\multiput(978.17,306.03)(6.000,-17.025){2}{\rule{0.400pt}{0.717pt}}
\multiput(985.59,284.08)(0.485,-1.408){11}{\rule{0.117pt}{1.186pt}}
\multiput(984.17,286.54)(7.000,-16.539){2}{\rule{0.400pt}{0.593pt}}
\multiput(992.59,264.88)(0.482,-1.485){9}{\rule{0.116pt}{1.233pt}}
\multiput(991.17,267.44)(6.000,-14.440){2}{\rule{0.400pt}{0.617pt}}
\multiput(998.59,248.43)(0.482,-1.304){9}{\rule{0.116pt}{1.100pt}}
\multiput(997.17,250.72)(6.000,-12.717){2}{\rule{0.400pt}{0.550pt}}
\multiput(1004.59,234.26)(0.482,-1.033){9}{\rule{0.116pt}{0.900pt}}
\multiput(1003.17,236.13)(6.000,-10.132){2}{\rule{0.400pt}{0.450pt}}
\multiput(1010.59,222.26)(0.482,-1.033){9}{\rule{0.116pt}{0.900pt}}
\multiput(1009.17,224.13)(6.000,-10.132){2}{\rule{0.400pt}{0.450pt}}
\multiput(1016.59,211.09)(0.482,-0.762){9}{\rule{0.116pt}{0.700pt}}
\multiput(1015.17,212.55)(6.000,-7.547){2}{\rule{0.400pt}{0.350pt}}
\multiput(1022.59,202.65)(0.482,-0.581){9}{\rule{0.116pt}{0.567pt}}
\multiput(1021.17,203.82)(6.000,-5.824){2}{\rule{0.400pt}{0.283pt}}
\multiput(1028.00,196.93)(0.491,-0.482){9}{\rule{0.500pt}{0.116pt}}
\multiput(1028.00,197.17)(4.962,-6.000){2}{\rule{0.250pt}{0.400pt}}
\multiput(1034.00,190.94)(0.774,-0.468){5}{\rule{0.700pt}{0.113pt}}
\multiput(1034.00,191.17)(4.547,-4.000){2}{\rule{0.350pt}{0.400pt}}
\multiput(1040.00,186.95)(1.132,-0.447){3}{\rule{0.900pt}{0.108pt}}
\multiput(1040.00,187.17)(4.132,-3.000){2}{\rule{0.450pt}{0.400pt}}
\put(1046,183.17){\rule{1.300pt}{0.400pt}}
\multiput(1046.00,184.17)(3.302,-2.000){2}{\rule{0.650pt}{0.400pt}}
\put(1052,181.67){\rule{1.445pt}{0.400pt}}
\multiput(1052.00,182.17)(3.000,-1.000){2}{\rule{0.723pt}{0.400pt}}
\put(1058,181.67){\rule{1.686pt}{0.400pt}}
\multiput(1058.00,181.17)(3.500,1.000){2}{\rule{0.843pt}{0.400pt}}
\put(1065,182.67){\rule{1.445pt}{0.400pt}}
\multiput(1065.00,182.17)(3.000,1.000){2}{\rule{0.723pt}{0.400pt}}
\put(1071,184.17){\rule{1.300pt}{0.400pt}}
\multiput(1071.00,183.17)(3.302,2.000){2}{\rule{0.650pt}{0.400pt}}
\put(1077,186.17){\rule{1.300pt}{0.400pt}}
\multiput(1077.00,185.17)(3.302,2.000){2}{\rule{0.650pt}{0.400pt}}
\multiput(1083.00,188.61)(1.132,0.447){3}{\rule{0.900pt}{0.108pt}}
\multiput(1083.00,187.17)(4.132,3.000){2}{\rule{0.450pt}{0.400pt}}
\multiput(1089.00,191.61)(1.132,0.447){3}{\rule{0.900pt}{0.108pt}}
\multiput(1089.00,190.17)(4.132,3.000){2}{\rule{0.450pt}{0.400pt}}
\multiput(1095.00,194.61)(1.132,0.447){3}{\rule{0.900pt}{0.108pt}}
\multiput(1095.00,193.17)(4.132,3.000){2}{\rule{0.450pt}{0.400pt}}
\multiput(1101.00,197.60)(0.774,0.468){5}{\rule{0.700pt}{0.113pt}}
\multiput(1101.00,196.17)(4.547,4.000){2}{\rule{0.350pt}{0.400pt}}
\multiput(1107.00,201.61)(1.132,0.447){3}{\rule{0.900pt}{0.108pt}}
\multiput(1107.00,200.17)(4.132,3.000){2}{\rule{0.450pt}{0.400pt}}
\put(1113,204.17){\rule{1.300pt}{0.400pt}}
\multiput(1113.00,203.17)(3.302,2.000){2}{\rule{0.650pt}{0.400pt}}
\multiput(1119.00,206.61)(1.132,0.447){3}{\rule{0.900pt}{0.108pt}}
\multiput(1119.00,205.17)(4.132,3.000){2}{\rule{0.450pt}{0.400pt}}
\put(1125,209.17){\rule{1.300pt}{0.400pt}}
\multiput(1125.00,208.17)(3.302,2.000){2}{\rule{0.650pt}{0.400pt}}
\put(1131,211.17){\rule{1.300pt}{0.400pt}}
\multiput(1131.00,210.17)(3.302,2.000){2}{\rule{0.650pt}{0.400pt}}
\put(1137,212.67){\rule{1.686pt}{0.400pt}}
\multiput(1137.00,212.17)(3.500,1.000){2}{\rule{0.843pt}{0.400pt}}
\put(1144,213.67){\rule{1.445pt}{0.400pt}}
\multiput(1144.00,213.17)(3.000,1.000){2}{\rule{0.723pt}{0.400pt}}
\put(493.0,215.0){\rule[-0.200pt]{2.891pt}{0.400pt}}
\put(1168,213.67){\rule{1.445pt}{0.400pt}}
\multiput(1168.00,214.17)(3.000,-1.000){2}{\rule{0.723pt}{0.400pt}}
\put(1174,212.17){\rule{1.300pt}{0.400pt}}
\multiput(1174.00,213.17)(3.302,-2.000){2}{\rule{0.650pt}{0.400pt}}
\put(1180,210.67){\rule{1.445pt}{0.400pt}}
\multiput(1180.00,211.17)(3.000,-1.000){2}{\rule{0.723pt}{0.400pt}}
\put(1186,209.17){\rule{1.300pt}{0.400pt}}
\multiput(1186.00,210.17)(3.302,-2.000){2}{\rule{0.650pt}{0.400pt}}
\put(1192,207.17){\rule{1.300pt}{0.400pt}}
\multiput(1192.00,208.17)(3.302,-2.000){2}{\rule{0.650pt}{0.400pt}}
\put(1198,205.17){\rule{1.300pt}{0.400pt}}
\multiput(1198.00,206.17)(3.302,-2.000){2}{\rule{0.650pt}{0.400pt}}
\multiput(1204.00,203.95)(1.132,-0.447){3}{\rule{0.900pt}{0.108pt}}
\multiput(1204.00,204.17)(4.132,-3.000){2}{\rule{0.450pt}{0.400pt}}
\put(1210,200.17){\rule{1.500pt}{0.400pt}}
\multiput(1210.00,201.17)(3.887,-2.000){2}{\rule{0.750pt}{0.400pt}}
\multiput(1217.00,198.95)(1.132,-0.447){3}{\rule{0.900pt}{0.108pt}}
\multiput(1217.00,199.17)(4.132,-3.000){2}{\rule{0.450pt}{0.400pt}}
\put(1223,195.17){\rule{1.300pt}{0.400pt}}
\multiput(1223.00,196.17)(3.302,-2.000){2}{\rule{0.650pt}{0.400pt}}
\put(1229,193.17){\rule{1.300pt}{0.400pt}}
\multiput(1229.00,194.17)(3.302,-2.000){2}{\rule{0.650pt}{0.400pt}}
\put(1235,191.17){\rule{1.300pt}{0.400pt}}
\multiput(1235.00,192.17)(3.302,-2.000){2}{\rule{0.650pt}{0.400pt}}
\put(1241,189.17){\rule{1.300pt}{0.400pt}}
\multiput(1241.00,190.17)(3.302,-2.000){2}{\rule{0.650pt}{0.400pt}}
\put(1247,187.17){\rule{1.300pt}{0.400pt}}
\multiput(1247.00,188.17)(3.302,-2.000){2}{\rule{0.650pt}{0.400pt}}
\put(1253,185.67){\rule{1.445pt}{0.400pt}}
\multiput(1253.00,186.17)(3.000,-1.000){2}{\rule{0.723pt}{0.400pt}}
\put(1259,184.17){\rule{1.300pt}{0.400pt}}
\multiput(1259.00,185.17)(3.302,-2.000){2}{\rule{0.650pt}{0.400pt}}
\put(1150.0,215.0){\rule[-0.200pt]{4.336pt}{0.400pt}}
\put(1271,182.67){\rule{1.445pt}{0.400pt}}
\multiput(1271.00,183.17)(3.000,-1.000){2}{\rule{0.723pt}{0.400pt}}
\put(1265.0,184.0){\rule[-0.200pt]{1.445pt}{0.400pt}}
\put(1283,181.67){\rule{1.445pt}{0.400pt}}
\multiput(1283.00,182.17)(3.000,-1.000){2}{\rule{0.723pt}{0.400pt}}
\put(1289,181.67){\rule{1.686pt}{0.400pt}}
\multiput(1289.00,181.17)(3.500,1.000){2}{\rule{0.843pt}{0.400pt}}
\put(1277.0,183.0){\rule[-0.200pt]{1.445pt}{0.400pt}}
\put(1308,182.67){\rule{1.445pt}{0.400pt}}
\multiput(1308.00,182.17)(3.000,1.000){2}{\rule{0.723pt}{0.400pt}}
\put(1314,183.67){\rule{1.445pt}{0.400pt}}
\multiput(1314.00,183.17)(3.000,1.000){2}{\rule{0.723pt}{0.400pt}}
\put(1320,184.67){\rule{1.445pt}{0.400pt}}
\multiput(1320.00,184.17)(3.000,1.000){2}{\rule{0.723pt}{0.400pt}}
\put(1326,185.67){\rule{1.445pt}{0.400pt}}
\multiput(1326.00,185.17)(3.000,1.000){2}{\rule{0.723pt}{0.400pt}}
\put(1332,186.67){\rule{1.445pt}{0.400pt}}
\multiput(1332.00,186.17)(3.000,1.000){2}{\rule{0.723pt}{0.400pt}}
\put(1338,187.67){\rule{1.445pt}{0.400pt}}
\multiput(1338.00,187.17)(3.000,1.000){2}{\rule{0.723pt}{0.400pt}}
\put(1344,188.67){\rule{1.445pt}{0.400pt}}
\multiput(1344.00,188.17)(3.000,1.000){2}{\rule{0.723pt}{0.400pt}}
\put(1350,189.67){\rule{1.445pt}{0.400pt}}
\multiput(1350.00,189.17)(3.000,1.000){2}{\rule{0.723pt}{0.400pt}}
\put(1356,190.67){\rule{1.445pt}{0.400pt}}
\multiput(1356.00,190.17)(3.000,1.000){2}{\rule{0.723pt}{0.400pt}}
\put(1296.0,183.0){\rule[-0.200pt]{2.891pt}{0.400pt}}
\put(1369,191.67){\rule{1.445pt}{0.400pt}}
\multiput(1369.00,191.17)(3.000,1.000){2}{\rule{0.723pt}{0.400pt}}
\put(1362.0,192.0){\rule[-0.200pt]{1.686pt}{0.400pt}}
\put(1381,192.67){\rule{1.445pt}{0.400pt}}
\multiput(1381.00,192.17)(3.000,1.000){2}{\rule{0.723pt}{0.400pt}}
\put(1375.0,193.0){\rule[-0.200pt]{1.445pt}{0.400pt}}
\put(1405,192.67){\rule{1.445pt}{0.400pt}}
\multiput(1405.00,193.17)(3.000,-1.000){2}{\rule{0.723pt}{0.400pt}}
\put(1387.0,194.0){\rule[-0.200pt]{4.336pt}{0.400pt}}
\put(1417,191.67){\rule{1.445pt}{0.400pt}}
\multiput(1417.00,192.17)(3.000,-1.000){2}{\rule{0.723pt}{0.400pt}}
\put(1411.0,193.0){\rule[-0.200pt]{1.445pt}{0.400pt}}
\put(1429,190.67){\rule{1.445pt}{0.400pt}}
\multiput(1429.00,191.17)(3.000,-1.000){2}{\rule{0.723pt}{0.400pt}}
\put(1423.0,192.0){\rule[-0.200pt]{1.445pt}{0.400pt}}
\end{picture}
